\documentclass[prl, twocolumn, showpacs, floatfix, letterpaper, groupedaddress]{revtex4}
\usepackage[dvips]{graphicx}
\begin{document}

\title{Non-equilibrium two-phase coexistence in a confined granular layer}
\author{Alexis Prevost}
\altaffiliation{Present Address: Laboratoire des Fluides
Organis\'es, CNRS-UMR 7125, Coll\`ege de France, 11 place Marcelin
Berthelot 75231 Paris cedex 05, FRANCE\\}
\author{Paul Melby}
\author{David A. Egolf}
\author{Jeffrey S. Urbach}
\altaffiliation{Corresponding Author}
\email[]{urbach@physics.georgetown.edu}
\affiliation{Department of Physics, Georgetown University, 37th \& O Streets, Washington DC
20057, USA}

\date{\today}


\begin{abstract}

We report the observation of the homogenous nucleation of crystals in a dense
layer of steel spheres confined between two horizontal plates vibrated
vertically. Above a critical vibration amplitude, two-layer crystals with square
symmetry were found to coexist in steady state with a surrounding granular
liquid.  By analogy to equilibrium hard sphere systems, the phase behavior may
be explained through entropy maximization. However, dramatic non-equilibrium
effects are present, including a significant difference in the granular
temperatures of the two phases.

\end{abstract}

\maketitle

Statistical mechanics provides a powerful formalism for predicting
the behavior of systems at or near equilibrium. Many natural
phenomena, however, occur far from equilibrium, and extensions of
the machinery of statistical mechanics to situations where
significant energy flows are present would have a wide range of
potential applications.  Some success has been achieved in
extending statistical mechanics to far-from-equilibrium
situations.  For example, techniques for calculating the relative
probabilities of different configurations have been successfully
developed for non-equilibrium steady states in a few restricted
situations \cite{onsager,derrida,egolf}. Also, effective
temperatures based on the fluctuations in non-equilibrium steady
states have been developed recently and, in some cases, have been
found to equilibrate across different fluctuating quantities
\cite{danna,makse,edwards,ono,gallavotti,goldburg,aumaitre}. To
generalize these initial successes to a broader spectrum of
phenomena and to develop insights into other facets of a
statistical theory of far-from-equilibrium systems, new model
systems for investigating non-equilibrium steady states must be
developed and studied.  Here we report our investigations of a
simple, far-from-equilibrium granular system that shows 
that some mechanisms that operate in equilibrium appear to persist
into situations far-from-equilibrium, whereas other basic tenets
of equilibrium statistical mechanics must be substantially
modified.

Granular materials are ubiquitous in nature and show a remarkable
range of non-equilibrium behavior \cite{jaeger,kadanoff,poschel}.
Dynamic steady-states, achieved when energy input from an external
source balances energy lost through inelastic collisions, provide
an ideal testing ground for extensions of equilibrium statistical
mechanics \cite{danna,makse,edwards,aumaitre}. In this Letter, we
report experimental measurements and computer simulations of the
dynamics of spherical particles confined between two horizontal
vibrating parallel plates. We observe a freezing transition from a
homogeneous, disordered liquid to an ordered solid with square
symmetry coexisting with a surrounding liquid. An essentially identical transition
is observed in confined hard-sphere colloidal suspensions in
equilibrium \cite{pansu1,pieranski,pansu2,schmidt1,schmidt2},
where it is driven by entropy maximization.  The presence in both the granular and the
colloidal system of a solid phase with the same 
unexpected symmetry which occurs under the same geometric constraints and
at similar
densities strongly suggests a common mechanism.  Unlike the equilibrium
system, however, we find that the coexisting phases have dramatically
different granular temperatures, demonstrating that the ``zeroth
law'' of thermodynamics is not followed by the granular
temperature. Taken together, these results show that the driving
mechanism behind an equilibrium phase transition may still operate
far from equilibrium, but that a thermodynamic theory must account
for the absence of equipartition in the kinetic energy of the
particles.

\begin{figure}[!ht]
\includegraphics[width=7cm]{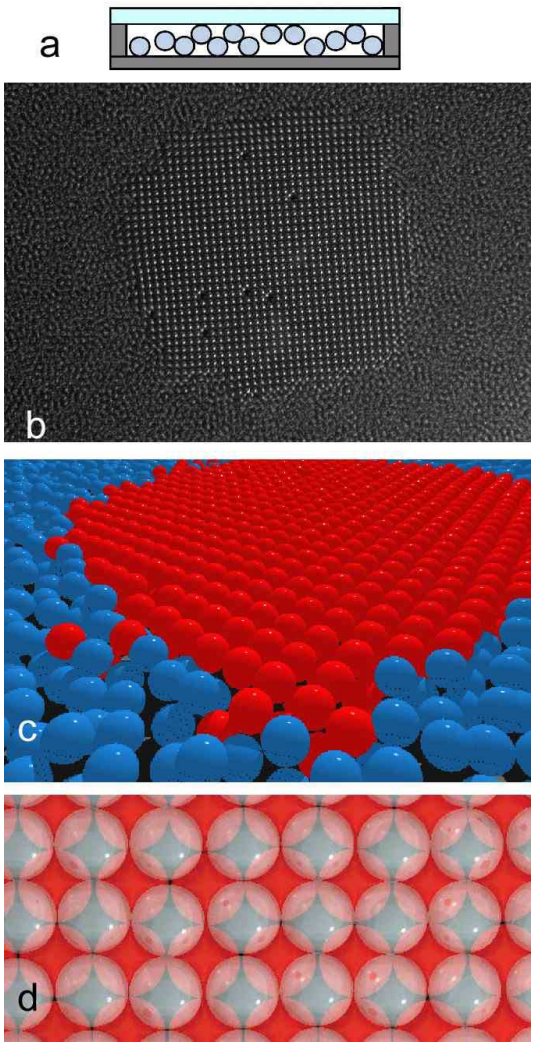}
\caption{ Two phase coexistence in steady state. (a) Side view of
experimental setup. The system is shaken vertically and imaged
from above. (b) Experiment: time-averaged image of ball positions.
Only the top layer of the crystal is visible and there are several
vacancies ($A=0.085~\sigma$, $\rho=0.9$, $\nu=80$~Hz, averaged
over 1~second.)  See also Supplemental Movie 1 \cite{movie}.
 (c, d) Simulation: 3D rendering of instantaneous ball
positions. In (c), balls in the crystal are colored red, balls in the
liquid are colored blue.  In (d) a close-up of the crystal is shown with the top layer transparent. ($A=0.13~\sigma$, $\rho=0.89$,
$\nu=60$~Hz, and 5,000 balls.) } \label{fig:coexistence}
\end{figure}

The granular system is sketched in Fig.~\ref{fig:coexistence}(a). Previous
studies which used a similar geometry at lower shaking amplitude
and lower density than what is presented here found a range of
complex non-equilibrium phenomena including inelastic collapse
\cite{olafsen1,olafsen2}, hexagonal ordering
\cite{olafsen1,pieranski2}, non-Gaussian velocity distributions
\cite{olafsen1,olafsen2,losert}, and velocity correlations
\cite{prevost}. In our experiment, stainless steel spheres of
diameter $\sigma=1.59$~mm were placed between a smooth anodized
aluminum plate and an 11~mm thick Plexiglas lid with an
anti-static coating.  A gap spacing of $1.75~\sigma$ between the
plate and the lid was maintained by circular rings of aluminum and
Mylar spacers. Using an electromagnetic shaker, the system was
driven sinusoidally in the vertical direction with frequency $\nu$
and amplitude $A$. The motion of the balls was imaged from above
using a high resolution camera (Pulnix TM1040). The results presented 
here were obtained
for densities $\rho=N/N_{max}$ ranging from 0.8 to 0.9, where $N$
is the number of balls in the system, and $N_{max}=11377$ is the
maximum number of balls that can fit in a single hexagonally
close-packed layer at rest on the bottom plate.  For modest
vibration amplitude, the system appears liquid-like. As the
vibration is increased, small independent unstable crystalline
structures form. Increasing the vibration amplitude further causes
the crystallites to become stable and eventually coalesce to form
a single crystal which coexists with a surrounding granular liquid
(see Fig. \ref{fig:coexistence}(b) and Supplemental Movie 1 \cite{movie} ) 
The crystal
consists of two layers, each with a square symmetry. The balls in
the second layer are above the centers of the squares formed by
the balls in the bottom layer. The crystals are not densely packed
and the balls constantly jitter around in the cage formed by their
neighbors and the confining plates. Rearrangements continually
occur at the interface, but the average size of the crystal does
not change as long as the shaking amplitude and frequency are held
constant.

In order to verify that the coexistence was not due to
non-uniformities in the experimental apparatus and to measure
quantities not readily accessible in the experiment, we performed
molecular dynamics simulations using a model that has accurately
reproduced many of the phenomena observed in a similar system
\cite{prevost,nie}. Ball-ball, ball-plate, and ball-lid
interactions are characterized by three forces: an elastic
restoring force, a dissipative normal force which produces a
velocity-independent coefficient of restitution, and a dissipative
tangential friction. Periodic boundary conditions in the
horizontal plane were used. The simulations reproduced the
phase coexistence (Fig. \ref{fig:coexistence}(c, d)) and all of the
general features of the observed phenomena, such as the existence
of a critical threshold to nucleation and evaporation.

Perhaps the most surprising aspect of the transition we observe,
the presence of a square symmetry instead of the hexagonal
ordering naively expected for hard-sphere interactions, appears
to be closely related to the phase behavior of solutions of
hard sphere colloidal particles at similarly high densities in similar confining
geometries \cite{pansu1,pieranski,pansu2}. For
hard spheres the equilibrium configuration is determined by
entropy maximization. For a range of gap spacings, including the
spacing used in our system, two square layers pack more
efficiently than two hexagonal layers, thereby maximizing the free
volume available for each particle and therefore the entropy of
the system \cite{schmidt1,schmidt2}. The observation of a transition that closely matches an equilibrium, entropy
driven phase transition suggests that a generalized
free energy functional might be found which describes the behavior
of some driven granular materials.

\begin{figure}[!ht]
\includegraphics[width=6cm]{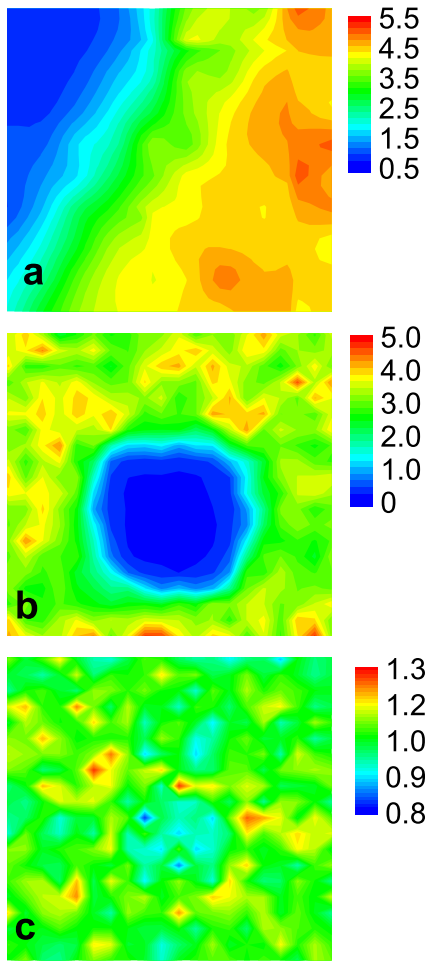}
\caption{ Temperature and pressure fields. (a) Temperature field
near the liquid-solid interface, measured in the experiment,
averaged over about 40 seconds. The crystal is in the upper left.
($\rho=0.85$, $\nu=60$~Hz, $A=0.154~\sigma$, displacements measured 
over 0.33 ms.)
(b), Temperature field and (c) pressure field from the simulation
shown in Fig. 1c., averaged over 2.5 seconds.  The crystal is in
the center. The pressure was calculated using the virial
expression, as described in Ref. \cite{rapaport}.}
\label{fig:temperature}
\end{figure}

Equilibrium two-phase coexistence requires that the two phases
have equal temperatures, pressures (apart from surface tension
corrections) and chemical potentials. Recent work has focused on
extending the concept of the chemical potential to non-equilibrium
coexistence in which the first two conditions are satisfied
\cite{baranyi,butler}. To test whether these two conditions are
met in this system, we measured the granular temperature
$T_g=<{\bf v}_i^2>$, where ${\bf v}_i$ is a horizontal component
of the rapidly fluctuating velocity of a particle.  In the experiment, particle
displacements were measured using the PIV technique described in Ref. \cite{prevost}.  Using the
method described in Ref. \cite{losert}, we verified that the temporal resolution was sufficient to accurately
measure instantaneous velocities.   In both the
experiments and simulations, we investigated whether the granular
temperature equilibrated to the same value in the two phases. As
shown in Figs. \ref{fig:temperature}(a) and
\ref{fig:temperature}(b), $T_g$ is dramatically lower in the
crystal than in the surrounding liquid, both in the experiments
and in the simulations. The
spontaneous separation into phases of different temperatures in a
homogeneous system of identical particles is a striking effect
that will have to be incorporated into models of non-equilibrium
phase coexistence.  It is somewhat reminiscent of `inelastic collapse'\cite{olafsen1},
but in that case the absence of any significant granular temperature in the solid
phase arises from the bistability of the ball-plate dynamics at low vibration amplitudes \cite{losert99,geminard03}.  The results described here are observed at high vibration amplitudes 
where there is continuous energy input from the plate into both coexisting phases.
The pressures of the two phases
calculated in the simulations have nearly the same value, but is slightly smaller in the solid phase (Fig. \ref{fig:temperature}(c)).  

\begin{figure}[!ht]
\includegraphics[width=8cm]{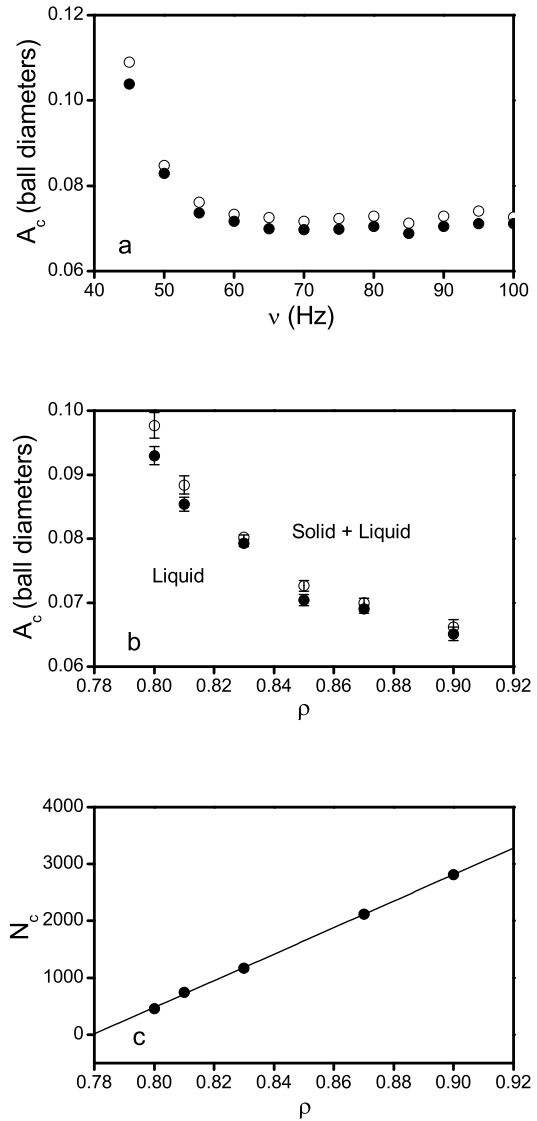}
\caption{(a) Frequency dependence of the critical amplitudes $A_C$
for nucleation (open circles) and evaporation (black disks) at
$\rho=0.85$. (b) High frequency plateau value of $A_C$ as a
function of $\rho$. (c) Average number of balls of the crystal,
$N_C$, as a function of $\rho$ ($\nu=60$~Hz, $A=0.145~\sigma$).
Each point represents the average of 10 measurements well separated in time . The solid line is a linear fit to the
points.} \label{fig:phase}
\end{figure}

To further study the properties of the phase coexistence, we
investigated the nucleation of the crystalline phase. Starting
with the vibration amplitude at a low value, we slowly increased
the intensity of shaking and measured the amplitude at which the crystalline phase
first nucleates.  This procedure was repeated for
several densities between $\rho=0.8$ and $\rho=0.9$, and for
frequencies between $\nu=45$~Hz and $\nu=100$~Hz. A typical curve
of the frequency dependence is displayed in Fig.
\ref{fig:phase}(a) for $\rho=0.85$. In addition to this
`nucleation line', we also determined an `evaporation line' by
slowly decreasing the amplitude until the crystal disappeared. For
$\nu$ greater than 60~Hz, we found that the critical amplitudes
were roughly independent of frequency. This high frequency
behavior was found at all densities, but the cutoff frequency
increased as the density increased. One possible explanation for a 
frequency-independent critical amplitude is that the vibration may effectively compresses the layer.  If the balls are moving slowly compared to the plate and lid, then they
will be mostly confined between the maximum plate height and minimum lid height.
  This increase in the density of the system favors nucleation of the crystal.
This frequency-independent behavior cannot persist to low frequencies, however, because
the acceleration, which is proportional to $\nu^2$, must be
significantly larger than that due to gravity for the balls to
have enough kinetic energy to reach the second layer. We used the
average value of the amplitude in the high frequency plateau
(60-100~Hz) to define a critical amplitude, $A_C$, and we
constructed a ``phase diagram'' of $A_C$ versus $\rho$ (Fig.
\ref{fig:phase}(b)). $A_C$ varies from roughly $0.06~\sigma$ at
$\rho=0.9$ to $0.1~\sigma$ at $\rho=0.8$.

We measured the dependence of the crystal size on the number of
particles in the system by analyzing the images to extract $N_C$,
the number of spheres in the crystal \cite{sizecrystal}.  As shown
in Fig. \ref{fig:phase}(c), $N_C$ varies linearly with $\rho$ and
extrapolates to zero at $\rho=0.78$. In steady state coexistence,
the edge of the crystal is ``in equilibrium'' with the surrounding
liquid of density $\rho_L$. Assuming that the densities of the
coexisting crystal and liquid are independent of the size of the
crystal, the area A occupied by the crystal of density $\rho_C$
should satisfy the relation A/A$_T$ $= (\rho - \rho_L)/(\rho_C -
\rho_L)$ where A$_T$ is the total surface area of the plate, so
that $N_C \propto (\rho - \rho_L)$. The value of $\rho_L$ found by
extrapolating to $N_C=0$ agrees with direct measurements of the
density of the granular liquid in the coexistence region.

No formalism
exists for incorporating the entropy into a predictive theory on non-equilibrium phase transitions, but our results indicate which parts of the
equilibrium framework need modification. The large difference in
the granular temperature of the coexisting phases demonstrates
that the ``zeroth law'' of thermodynamics is not satisfied by the
granular temperature. An effective temperature that does meet this
requirement is probably a necessary ingredient of a quantitative
theory of the phase coexistence.  By comparing the system described in this
Letter with the analogous and well understood equilibrium system, new approaches for incorporating the effects of forcing and dissipation into a
statistical mechanics of non-equilibrium phase transitions can be developed and tested.

\begin{acknowledgements}
This work was supported by The National Science Foundation with grants DMR-9875529 and DMR-0094178
and by NASA under award number NNC04GA63G.  D.A.E. is also
supported as an Alfred P. Sloan Research Fellow.
\end{acknowledgements}


\begin{thebibliography}{99}
\bibitem{sizecrystal} $N_C$ is taken as twice the number of balls in the top layer of the crystal,
averaged over several snapshots well separated in time. An algorithm based on a four-fold symmetry
order parameter and related criteria was employed in determining which balls were in the crystal.
\end{thebibliography}
\end{document}